\documentclass[%
 reprint, 
 amsmath,amssymb,
 aps,
pra,
]{revtex4-2}
\usepackage[T1]{fontenc}
\usepackage[utf8]{inputenc}
\usepackage[english]{babel} 
\usepackage{graphicx}
\usepackage{dcolumn}
\usepackage{bm}
\usepackage{braket}
\usepackage{tabularx}
\usepackage{amsmath}
\usepackage{float} 
\usepackage{bbm}
\DeclareMathAlphabet{\mathbbold}{U}{bbold}{m}{n} 
\usepackage{subfig}

\begin{document}

\preprint{APS/123-QED}

\title{Continuous variables graph states shaped as complex networks: \\optimization and manipulation.}

\author{Francesca Sansavini}
 \altaffiliation[]{francesca.sansavini@lkb.upmc.fr}
\author{Valentina Parigi}%
 \email{valentina.parigi@lkb.upmc.fr}
\affiliation{%
Laboratoire Kastler Brossel, Sorbonne Universit\'{e}, CNRS, ENS-PSL Research University, Coll\`{e}ge de France\\ 4 place Jussieu, F-75252 Paris, France; 
}%

\date{\today}

\begin{abstract}
Complex networks structures have been extensively used for describing complex natural and technological systems, like the Internet or social networks. 
More recently complex network theory has been applied to quantum systems, where complex network topologies may emerge in multiparty quantum states and quantum algorithms have been studied in complex graph structures. In this work we study multimode Continuous Variables entangled states, named cluster states, where the entanglement structure is arranged in typical real-world complex networks shapes.
Cluster states are a resource for measurement-based quantum information protocols, where the quality of a cluster is assessed in term of the minimal amount of noise it introduces in the computation. We study optimal graph states that can be obtained with experimentally realistic quantum resources, when optimized via analytical procedure. We show that denser and regular graphs allow for better optimization. In the spirit of quantum routing we also show the reshaping of entanglement connections in small networks via linear optics operations based on numerical optimization.
\end{abstract}

\maketitle


\section{Introduction}
 In the last decades network theory has provided a natural framework for describing complex natural, social and technological structures \cite{Al02, Liu16}. Recurrent types of complex networks, like the free-scale networks, have been recovered in phenomena at different scales, where the functionality of the systems seems to be closely related to their structure \cite{Al00, Gao16}. 
More recently complex networks have gained attention in the quantum realm, where several theoretical works \cite{Bia15,Bia19,Ac07,Fac14,Pap13,Cab18} show that complex structures may play a role in quantum information technologies. It is clear that, as quantum architectures are reaching larger scales, their internal arrangement starts to play a significant role in their functionality. Moreover network structures are clearly at the base of quantum communication protocols \cite{Kimb08}, but also appear in particular kind of multiparty entangled states that allow for measurement based quantum computing (MBQC) protocols \cite{Bri1W}.  \\
Recently, new records have been established in superconductor-\cite{Aru19,IBM, Rea18}  and Rydberg-\cite{Ber17, Bar18}   based technologies, and extremely large entangled states have been generated in the optical domain \cite{Asa19,Cai, Che14,Yosh16}. We can describe superconducting and Rydberg platforms as networks of interacting qubits, where quantum information is encoded in Discrete Variables (DV) states. Differently, in the optical domain, networks of entangled optical modes are deterministically generated and measurement-based quantum protocols can be implemented via a Continuous Variables (CV) encoding \cite{Gu}.  The control on the reconfigurability of the cited networks is gradually increasing, and in particular, in the CV case, it is possible to have totally reconfigurable networks (with all-to-all connections) mimicking complex networks structures \cite{Nok18}.\\
Here we study CV entangled states, named CV graph states or CV cluster states, arranged with shapes that are typical of real-world complex networks, in order to investigate their properties for quantum computing or quantum networking (communication) protocols. \\
CV graphs have been introduced in the context of measurement-based quantum computing exploiting CV resources \cite{Men06,Gu}.While usually the name cluster is used when the graph shape allows for universal quantum computing, in this work we will use the terms cluster state and graph state as synonyms. \\ In CV-quantum computing quantum information is encoded in variables that can take a continuum spectrum of values: in optics these variables are the quadrature of the electromagnetic field (called amplitude and phase quadrature) and they correspond to the position $q$ and momentum $p$ variables of the harmonic oscillator which represents the behavior of a single mode of the field.  
Ideal cluster states, which involve perfect correlations between quadratures of different optical modes, cannot be reached experimentally, as they would require an infinite amount of energy (squeezing) to be established. So experimental clusters always involve a poorer degree of correlations, which causes errors and noise in computation. Given experimentally accessible resources, analytical optimization protocols have been proposed to choose state manipulations that better distribute the correlations in order to minimize errors \cite{Giulia1}, then improving the quality of the cluster.  In this work we apply the optimization procedure to complex cluster shapes and we find that denser and more regular network shapes allow for better quality. \\
As the classical internet can be described via a complex network model, it is worth to consider complex network shapes in quantum communication protocols. In particular we are interested in quantum routing, i.e. the establishment of a direct entanglement link between two arbitrary nodes of a network in order to have an exploitable channel for quantum teleportation. Quantum routing protocols have been mainly studied in the DV approach \cite{Ha19,Pir18}, while here we study the particular setting of CV quantum resources. In particular, instead of considering a set of initially disconnected nodes subjected to routing via local operations, we consider a network of CV entanglement correlations, which can be easily produced with the current technology \cite{Cai, Che14,Yosh16}, and then we reconfigure it via partially-local and easy operations (linear optics). The optimal operations are obtained via numerical optimization. We found solutions for small networks of different geometry which are shared between two parties that are allowed to act via local linear optics operations in their set of nodes.

\section{Results}
\subsection{Background: Cluster states and complex networks}
Ideal cluster states are built starting from a set of modes of light, which are placed in the zero-momentum eigenstate $\vert 0\rangle_p$, i.e. infinitely squeezed vacuum states along the momentum quadrature $p$. 

Then entangling $C_Z= \exp(\imath \hat{q}_i \otimes \hat{q}_j)$ gates are applied between couples of modes $i$ and $j$,  according to a given configuration, which can be represented by a graph.  
A graph is defined by a set $\{\bf{\mathcal{V}},\bf{\mathcal{E}}\}$ where $\bf{\mathcal{V}}$ is the set of nodes (vertices) and $\bf{\mathcal{E}}$ are the edges. In the graphical representation of the cluster the nodes represent different modes of light, while the edges between the nodes are the entanglement correlations given by the application of the  $C_Z$ gates.  The graph is characterized by the adjacency matrix $\textbf{V}$ whose elements $V_{ij}$ are  set to $1$ when an entangling gate has been applied between the  two nodes $i$ and $j$, and $0$ otherwise. \\
An ideal cluster state of $N$ modes with adjacency matrix $\bf{\textbf{V}}$ is given by 
\begin{equation}
\ket{\Psi_C} =\prod_{1\leq i< j<N} \exp(\imath V_{ij}  q_i \otimes q_j) \vert 0\rangle_p^{\bigotimes^N}
\end{equation}
Cluster states are characterized by a particular set of operators, called \textit{nullifiers}, that read
\begin{equation}
\hat{\delta}_i=\hat{p}_i-\sum_{j\in \mathcal{N}(i)}\hat{q}_j, \quad\text{such that}\quad\hat{\delta}_i\ket{\Psi_C}=0
\end{equation}
where $\hat{\delta}_i$ and $\mathcal{N}(i)$ denote respectively the nullifier and the set of the nearest neighbors of the i-th node.
Cluster states are eigenstates of the nullifier with zero eigenvalue, so for an ideal cluster the following condition holds: $\Delta^2\delta_i=0$.
In addition, being Gaussian states, cluster states are fully described by mean values of quadratures and covariance matrices~\cite{Weed1}. 
By defining the vector $\hat{\textbf{X}} =\{\hat{q}_1,...,\hat{q}_N,\hat{p}_1,...,\hat{p}_N \}$ cluster states are identified by mean values $\langle \hat{X}_i \rangle =0$ and covariance matrix $\sigma$ with elements $\sigma_{i,j}=(1/2)\langle (\hat{X}_i \hat{X}_j+\hat{X}_j \hat{X}_i)\rangle$.

As already said, in a realistic situation, CV cluster states are always imperfect, due to the impossibility of reaching an infinite amount of squeezing  to obtain  the zero momentum eigenstate.  Approximated clusters, when used in measurement -based quantum computing protocols, at each step of the computation, introduce a certain amount of noise that can be quantified with the variance of the nullifiers of the cluster \cite{Gu}. Therefore, the smaller the value of $\Delta^2\delta_i$ is  for every node,  the better the quality of cluster is.

Instead of acting with $C_Z$ gate on multimode squeezed vacuum states, as in the original formulation, cluster states can be implemented  via linear optics transformations on the multimode squeezed vacuum state  \cite{Bra1, Loock1}. This is the implementation which is pursued in experimental setups \cite{Asa19,Yosh16,Che14,Cai}, as it is easier and less costly than  the implementation of $C_Z$, which requires online squeezing. 

Linear optics manipulations are described by unitary transformations $U$ on the creation and annihilation operators of the different light modes, so that the annihilation operator of the mode $i$ is transformed as  $\hat{a}_i\to \sum_j U_{ji}\hat{a}_i$. This corresponds to orthogonal symplectic transformations $S$ acting on the quadratures, which transform the covariance matrix as $\sigma \rightarrow  S \sigma S^T$.

For implementing a cluster with adjacency matrix $\textbf{V}$, we can apply the following unitary operation 
\begin{equation}\label{uniforV}
U=( \mathrm{I}+\imath \textbf{V})(\textbf{V}^2+\mathrm{I})^{-\frac{1}{2}} O
\end{equation}
where $\mathrm{I}$ is the identity matrix and $O$ is an arbitrary real orthogonal matrix. This orthogonal matrix provides supplementary $N(N-1/2)$ degrees of freedom, that can be exploited to optimize given properties of the cluster  \cite{Giulia1}. In particular we can optimize given properties of the nullifiers via an analytical protocol ~\cite{Adrien}, with the aim, for example,  of  reducing their variances, hence  improving the quality of the cluster.  \\
In this work we investigate, for the first time, clusters whose graphical representation $\{\bf{\mathcal{V}},\bf{\mathcal{E}}\}$  corresponds to three different models of complex networks \cite{Al02}: the Barabási-Albert \cite{Barabasi1}, the Erdős–Rényi \cite{WS1} and the Watts-Strogatz \cite{ER1}.    These models have been developed in graph theory in order to reproduce the behaviors of complex systems. The Erdős–Rényi has been the first considered model for complex networks. It is based on random graph: given a fixed number of nodes $n$, two nodes have a probability $p_{ER}$ to be connected by an edge.  The model generates graphs with approximately $p_{ER} n (n-1)/2$ edges which are distributed randomly.  Most nodes have a comparable node degree $k$, defined as the number of edges of the given node. \\
Later, by looking at the degree distribution of many real-world networks, it has been clear that complex networks go beyond the random graph model, as many networks (like Internet or the WWW)  are characterized by a power-law distribution of the degree.  So new models, as the  Barabási-Albert, have been developed in order to reproduce the structure of this networks, also known as scale-free networks.  Barabási-Albert networks grow from a small number of nodes according to preferential attachment, i.e. the fact that new nodes attach preferentially to nodes that already have a high degree (high number of links). In the growth process the parameter $m_{BA}$ specifies the number of links coming with the new node. Hubs, i.e. heavily linked nodes, arise spontaneously in this model. \\
Watts-Strogatz networks lie between regular and random graphs and exhibits small-world properties typical of social networks.  They are built from a regular network by rewiring its edges, to increase randomness.  From a ring lattice with $n$ nodes and $k$ edges per node, which connect it to the closest neighbors, each edge is randomly rewired with   probability $p_{WS}$.  The rewiring parameter $p_{WS}$ spans from 0, the regular graph, to 1, which corresponds to a totally random graph.
Even if we are going to study graphs with small number of nodes compared to the one for which these models have been developed, we can assess a different behavior for the corresponding graph states in the three classes.\\
 In the following we will see how, starting from a given set of modes with finite squeezing values, we can optimize the quality of the clusters corresponding to the these complex networks models,  and  how the result of the optimization depends on the  parameters of the analyzed model.

\subsection{Improving the overall quality of a complex cluster}
We test here cluster states with the structure of the three models defined above: Barabási–Albert (BA), Erdős–Rényi (ER) and Watts-Strogatz (WS), with different characterizing parameters $m_{BA}$, $p_{ER}$ and $p_{WS}$.

\begin{figure}[h]
\centering
\subfloat[][\emph{Barabási-Albert model with $m_{BA}=2$, with a maximum node degree of k=22}.]  
 {\includegraphics[width=0.4\columnwidth]{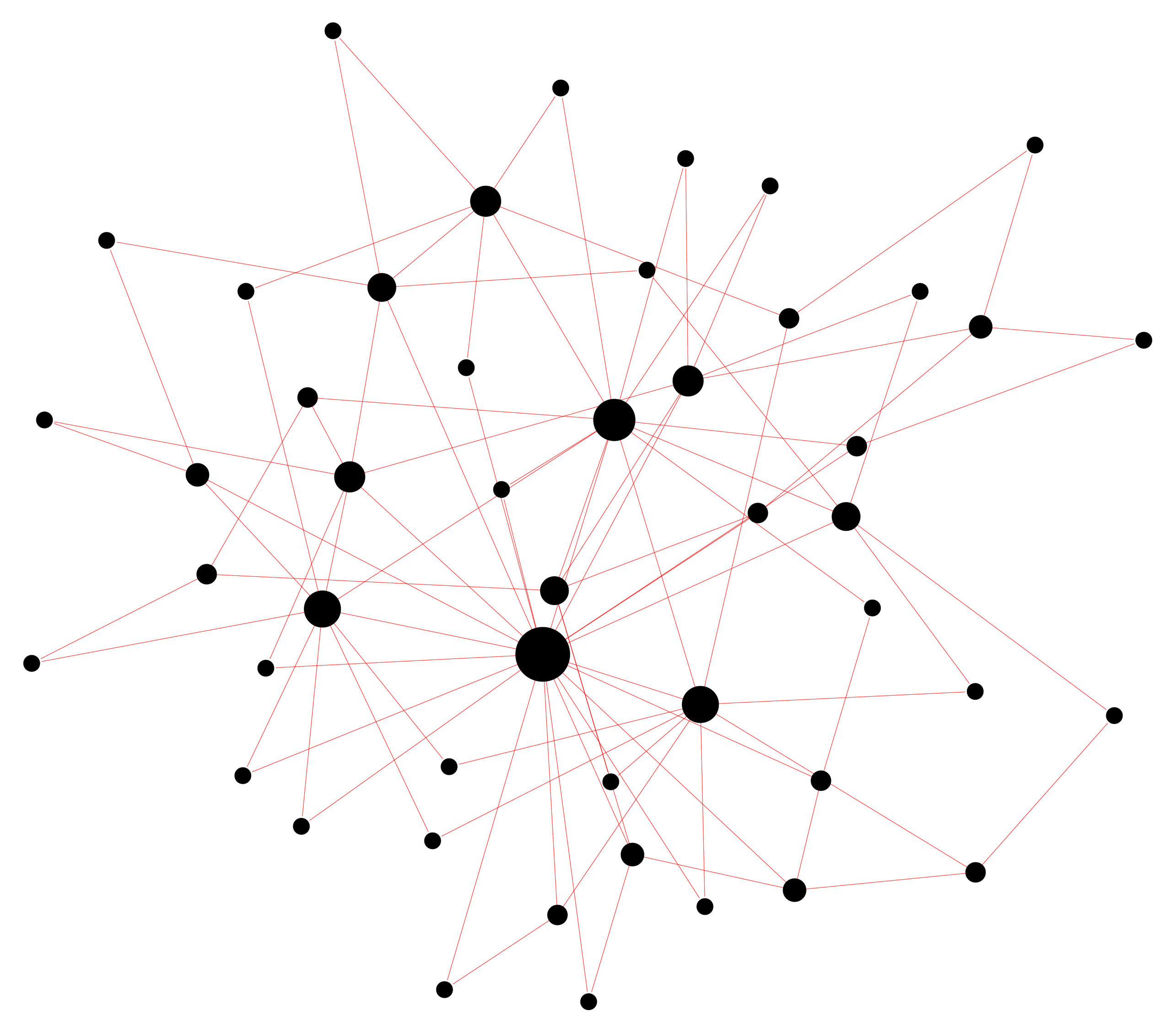}\label{BAcomp}} \quad
 \subfloat[][\emph{Erdős–Rényi model with $p_{ER}=4/49$, with a maximum node degree of k=8}.]
   {\includegraphics[width=0.4\columnwidth]{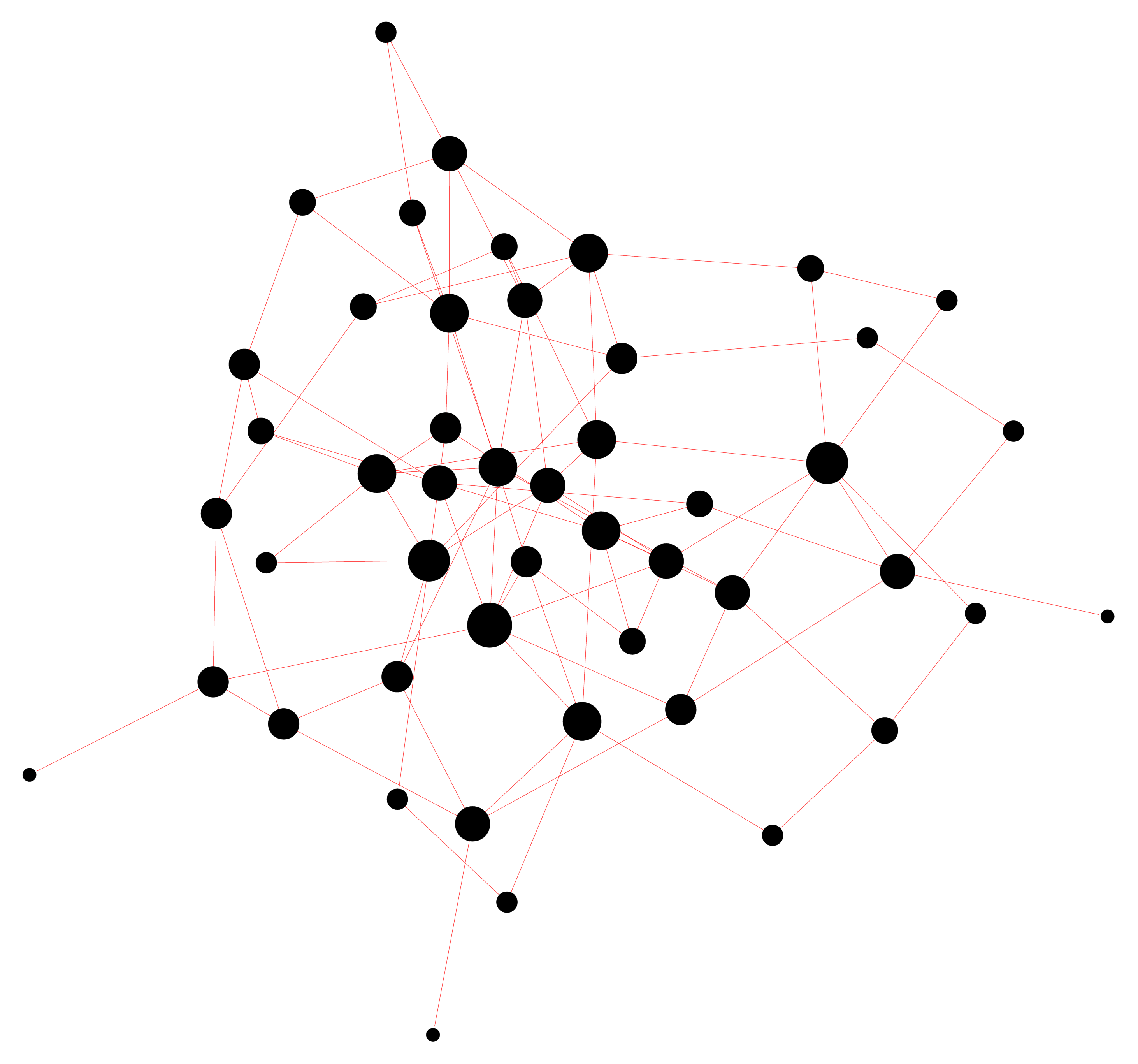}\label{ERcomp}} \quad
\caption{Comparison between two models of complex networks, both with an average degree of $\langle k \rangle \sim 3.9$. The size of the dots increases with the number of links.}
\label{compgraph}
\end{figure}

In Figure~\ref{compgraph}  the difference between a scale-free network and a random network with a comparable average degree is shown. As we see in Figure~\ref{BAcomp}, the Barabási-Albert network exhibits highly connected nodes (\textit{hubs}) that are absent in the Erdős–Rényi model (Figure~\ref{ERcomp}). 

\begin{figure}[h]
\centering
  \includegraphics[width=\columnwidth]{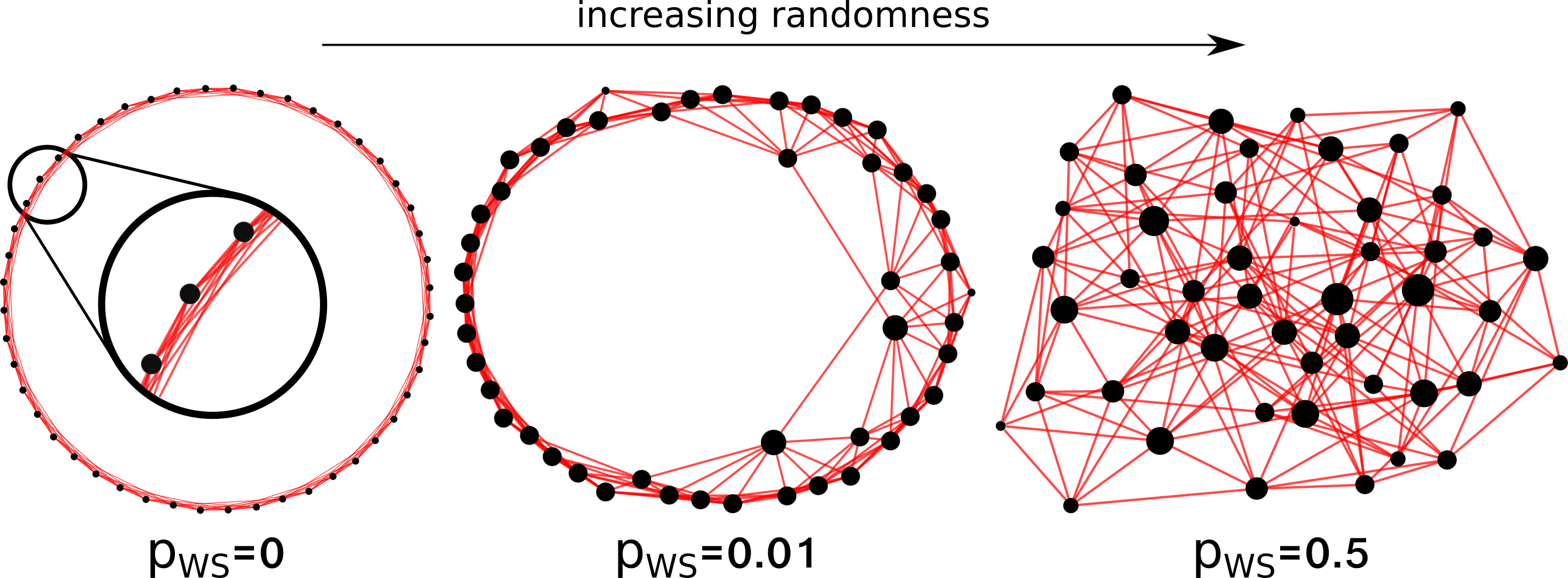}
\caption{Rewiring of a regular 48-node network for the construction of a ``small world'' network as shown in~\cite{WS1}.}
\label{wsgraphs}
\end{figure}

\begin{figure}[h]
\centering
\subfloat[][\emph{Implementation of a cluster via a linear optics transformation acting on a series of squeezed modes}.]  
 {\includegraphics[width=0.45\textwidth]{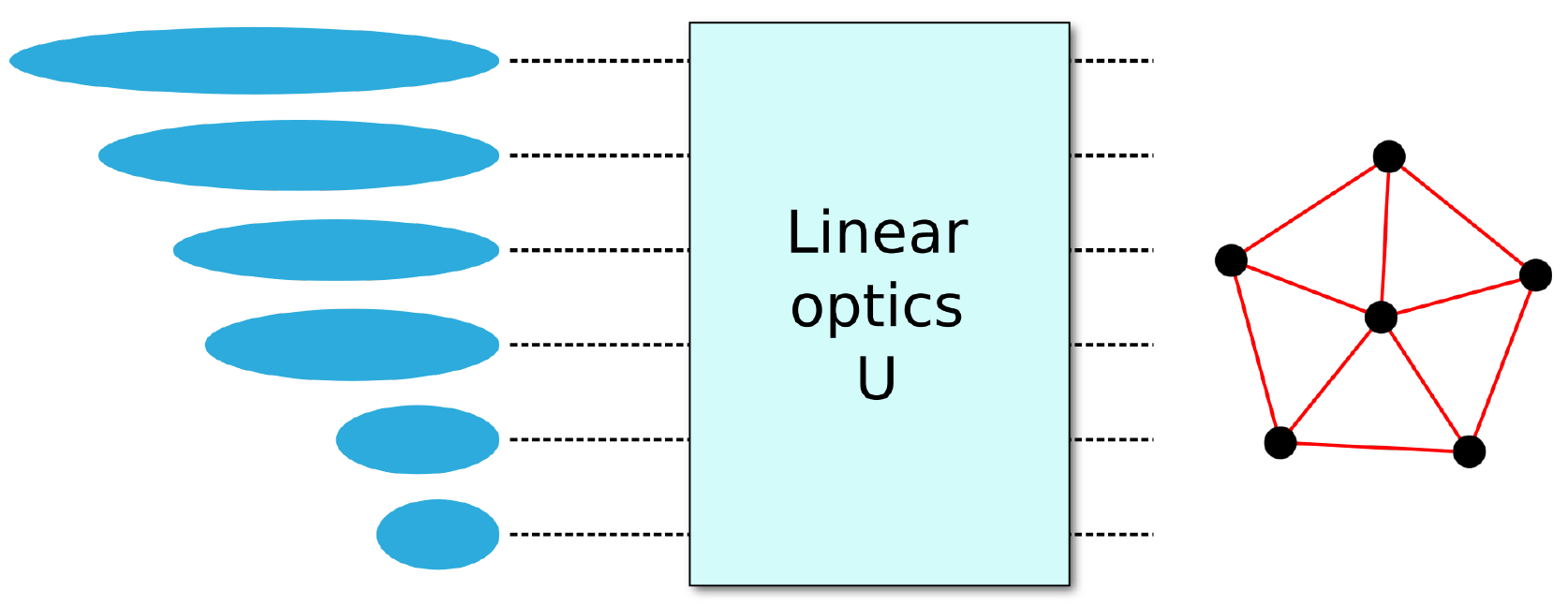}\label{linop}\label{linoptcluster}} \quad
 \subfloat[][\emph{List of realistic squeezing values of the input modes for the implementation of a 48-mode cluster}.]
   {\includegraphics[width=0.45\textwidth]{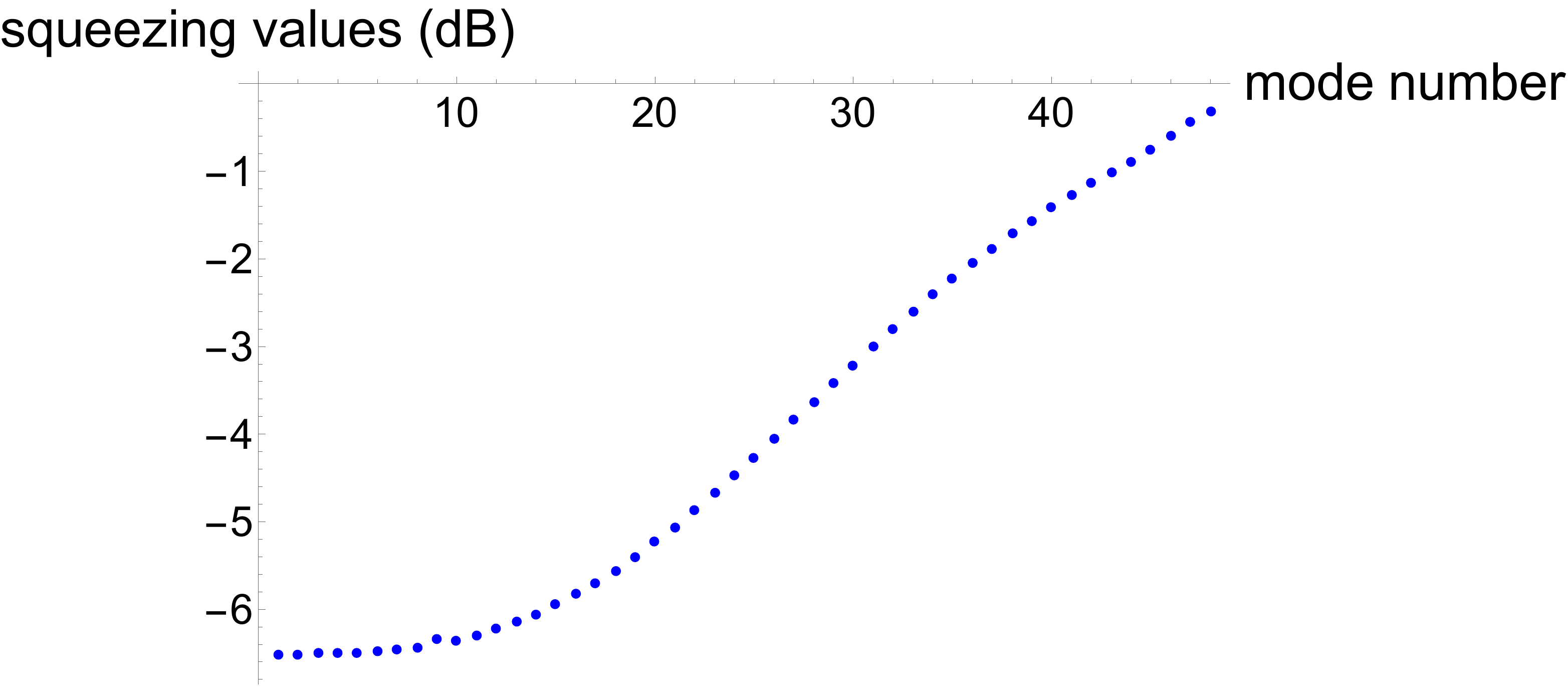}\label{listsqzval}} \quad
\caption{Realistic implementation of complex cluster states.}
\label{linoptexp}
\end{figure}

We consider at first the implementation of 48-mode complex clusters.
As explained above, the clusters are implemented from a 48-mode squeezed vacuum via the application of the linear optics unitary of Eq.~\eqref{uniforV} corresponding to the graphical shape we want to reach \cite{Giulia1}. The scheme is presented in Fig \ref{linop}. In order to pursue realistic implementations we consider the list of squeezing values presented in Fig.~\ref{listsqzval}, which corresponds to a series of  faithful values that can be obtained via the Schmidt decomposition of parametric process Hamiltonian of the experiment described in~\cite{Cai,PhDCai}. The obtained clusters are optimized, acting on the parameters of the arbitrary orthogonal matrix $O$ of Eq.~\eqref{uniforV}, by minimizing the fitness function $f(\Delta^2\delta_i)=\sum_i \Delta^2\bar{\delta}_i$, where $\Delta^2\bar{\delta}_i$ denotes the i-th nullifier normalized with the vacuum noise.  

As BA, ER and WS are statistical models for graph families, we implement $N=100$ graphs for each model with a given set of parameters. We define the average quality of a single graph $j$ as $\mu_j$ and the average quality for a set of graphs as $\mu$, as follows:
\begin{equation}
\label{mukappa}
\mu_j=\frac{1}{48}\sum_{i}\Delta^2\bar{\delta}_{i,j}\text{  }; \quad \mu=\langle \mu_j \rangle
\end{equation}
where $\Delta^2\bar{\delta}_{i,j}$ identifies variance of the $i$-th nullifier of the $j$-th graph. We will indicate as $\sigma$ the standard deviation of the $\mu_j$ values. This way we can average out the fluctuations due to the randomness of the complex shape.

\begin{table}[H]
\centering
\subfloat[][Barabási–Albert]{
\begin{ruledtabular}
\begin{tabular}{cccc}
$m_{BA}$		&$\mu$ (dB)	&$[\mu\pm\sigma]$ (dB)			&$\langle k \rangle$\\	
\colrule
1			&-4.70				&$[-4.73,-4.67]$		&1.96\\
5			&-5.55				&$[-5.58, -5.53]$		&9.38\\
10			&-5.82				&$[-5.84, -5.80]$		&17.71\\
20			&-6.15				&$[-6,16, -6.14]$		&31.25\\
47			&-6.33				&$[-6.33, -6.33]$		&47\\

\end{tabular}\end{ruledtabular}	\label{BAaaa}}\quad
\subfloat[][Erdős–Rényi]{
\begin{ruledtabular}
\begin{tabular}{cccc}
$p_{ER}$		&$\mu$ (dB)			&$[\mu\pm\sigma]$ (dB)	&$\overline{ \langle k_i \rangle}$\\
\colrule
0.2			&-5.50				&$[-5.54,-5.46]$		&9.35\\
0.4			&-5.80				&$[-5.83, -5.76]$		&18.83\\
0.6			&-6.02				&$[-6.04, -6.00]$		&28.29\\
0.8			&-6.22				&$[-6.23, -6.21]$		&37.58\\
1			&-6.33				&$[-6.33, -6.33]$		&47\\

\end{tabular}\end{ruledtabular}\label{ERaaa}}	
\caption{Mean $\mu$ and standard deviation $\sigma$ of the values $\mu_j$ of Eq.~\eqref{mukappa} evaluated on $N=100$ Barabási–Albert (a) and Erdős–Rényi (b) graphs with different characterizing parameters and consequently different average degrees $\langle k \rangle$, optimized using the function $f(\Delta^2\delta_i)=\sum_i \Delta^2\bar{\delta}_i$. Without the optimization protocol, $\mu$ takes the value of $-3.48$ dB for the Erdős–Rényi model, independently of the value of $p_{ER}$, and it oscillates between $-3.48$ dB and $-3.72$ dB for the Barabási–Albert model.}
\label{TableBA-ER}
\end{table}

As we can see from Table~\ref{TableBA-ER}, the implementation of quantum complex networks following the Barabási-Albert model or the Erdős–Rényi model shows that the quality of the cluster increases when the average number of edges per node, the average degree $\langle k \rangle$, increases. The average degree $\langle k \rangle $ can be raised by increasing the parameters $m_{BA}$ and $p_{ER}$, for the BA and the ER model respectively.
The results clearly show that the quality of the cluster $\mu$ increases with these parameters, until the limiting case of the fully connected graph ($m_{BA}=47$ for the Barabási-Albert model and $p_{ER}=1$ for the Erdős–Rényi model) is reached.  

\begin{table}[h]
\centering
\subfloat[][$\langle k \rangle=4$]{
\begin{ruledtabular}
\begin{tabular}{ccc}

$p_{WS}$		&$\mu$ (dB)	&$[\mu\pm\sigma]$ (dB)	\\	
\colrule
0			&-5.19				&$[-5.19, -5.19]$\\
0.1			&-5.16				&$[-5.17, -5.14]$\\
0.4			&-5.10				&$[-5.12, -5.07]$\\
0.7			&-5.09				&$[-5.11, -5.07]$\\
1			&-5.09				&$[-5.12, -5.06]$\\

\end{tabular}\end{ruledtabular}\label{kappa4}}\quad	
\subfloat[][$\langle k \rangle=8$]{\begin{ruledtabular}\begin{tabular}{ccc}
$p_{WS}$		&$\mu$ (dB)	&$[\mu\pm\sigma]$ (dB)	\\	
\colrule
0			&-5.79				&$[-5.79, -5.79]$\\
0.1		&-5.69				&$[-5.71, -5.66]$\\
0.4		&-5.49				&$[-5.51, -5.46]$\\
0.7		&-5.43				&$[-5.46, -5.40]$\\
1			&-5.43				&$[-5.46, -5.41]$\\
\end{tabular}\end{ruledtabular}\label{kappa8}}
\caption{Mean $\mu$ and standard deviation $\sigma$ of the values $\mu_j$ of Eq.~\eqref{mukappa} evaluated on $N=100$ Watts-Strogatz graphs with different parameter $p_{WS}$ and different $\langle k \rangle$, optimized using the function $f(\Delta^2\delta_i)=\sum_i \Delta^2\bar{\delta}_i$. Without the optimization protocol, $\mu$ takes the value of $-3.48$ dB, independently of the value of $p_{WS}$ or $\langle k \rangle$.}
\label{WSktable}
\end{table}

\begin{figure}[h]
\centering
\includegraphics[width=\columnwidth]{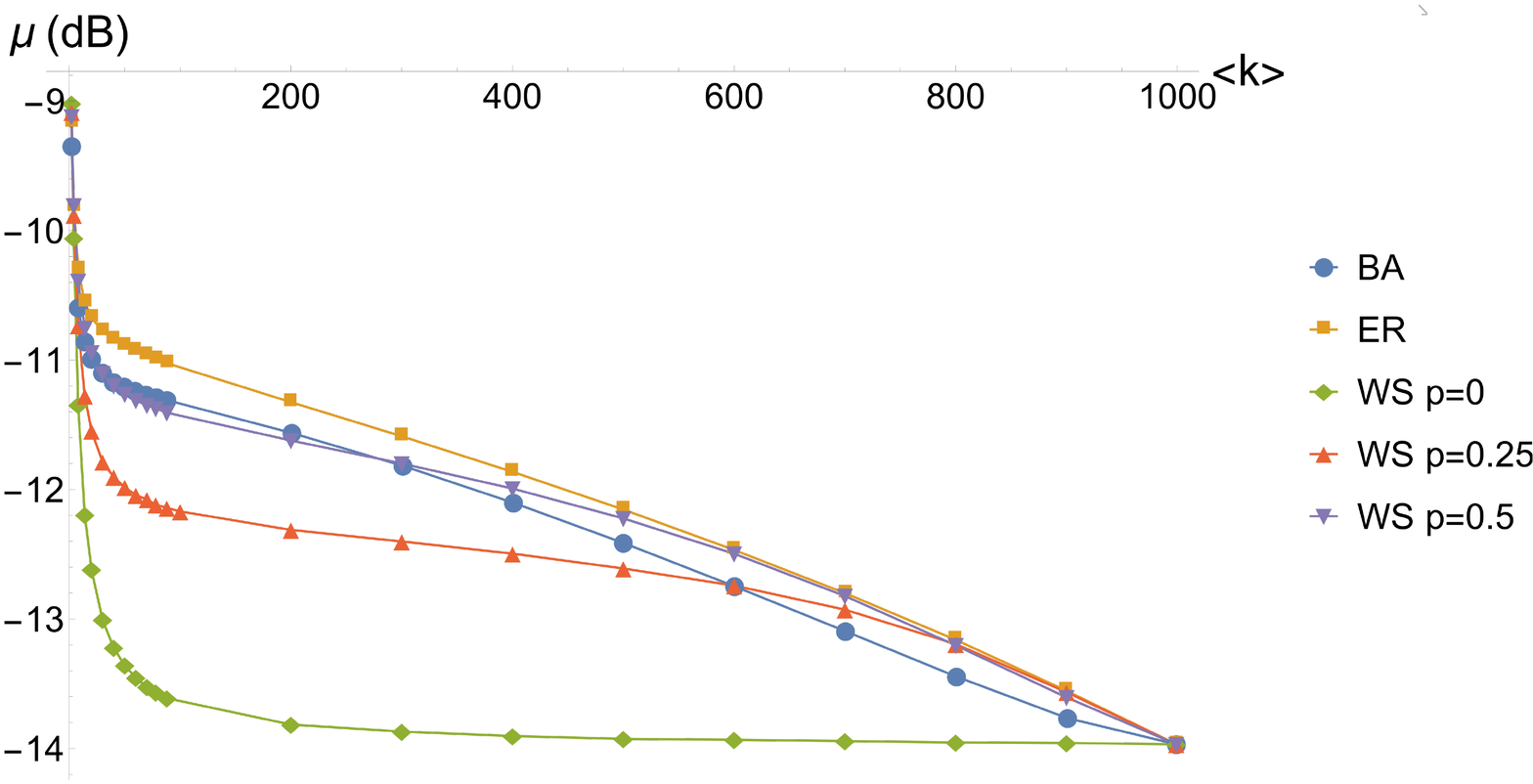}
\caption{Plot of the mean squeezing value of the nullifiers of the cluster as a function of its average degree $\langle k \rangle$ for the different topologies of complex graphs. In the legend, ``BA''=Barabási-Albert, ``ER''=Erdős–Rényi, ``WS p=0'' = Watts-Strogatz with $p_{WS}=0$, ``WS p=0.25'' = Watts-Strogatz with $p_{WS}=0.25$ and ``WS p=0.5'' = Watts-Strogatz with $p_{WS}=0.5$.}
\label{comparison}
\end{figure}

The Watts-Strogatz cluster confirms that the quality of the cluster increases for larger value of $\langle k \rangle$,  as we can see by comparing Table~\ref{WSktable}a and ~\ref{WSktable}b, but it also shows a peculiar behaviour that depends on $p_{WS}$. As already said, $p_{WS}$ is a rewiring parameter that can be varied from $0$ to $1$ in order to tune the randomness of the graph without changing the average degree, as the number of edges is not changed by the rewiring. From the Tables in~\ref{WSktable} it is clear that the quality of the cluster is reduced when $p_{WS}$ approaches 1, so that regular graphs states are optimized better than random graphs.  \\
The results we obtained on the 48-node graphs we analysed show a dependence between the quality of the optimization and the topology of the graph and in particular its dependence on the average degree $\langle k \rangle$. In order to reduce the influence of the finite size in the network models we have been using, we repeat the procedure for larger networks. We optimize the mean value $\mu$ of nullifiers squeezing (Eq.~\eqref{mukappa}) of a set of 10 complex graphs of 1000 nodes for a certain complex network model with given parameters. The optimization is reiterated by varying the average degree $\langle k \rangle$ and by varying the model. 
The initial list of 1000 squeezing values is obtained from a pseudorandom number generator that created uniformly distributed random numbers in the range $[-14, -3]$. The results are reported in Fig.~\ref{comparison}.

The data show that the regular cluster (Watts-Strogatz model with $p_{WS}=0$) converges very fast to its optimal overall quality. Indeed, as we expected from the 48-node cluster analysis, we see that, for a fixed $\langle k \rangle$, increasing the regularity, via the $p_{WS}$ parameter, results in a better quality of the optimized cluster and in a faster convergence, as we can see comparing the curves for the Watts-Strogatz model with $p_{WS} = 0$,  $p_{WS}=0.25$ and $p_{WS}=0.5$. On the other hand, the difference among the Barabási-Albert, Erdős–Rényi and Watts-Strogatz $p_{WS}=0.5$ complex graphs is less significant. Nevertheless, the convergence of the Watts-Strogatz follows a different behaviour, being closer to the Barabási-Albert for lower $\langle k\rangle$ and converging to the behaviour of the Erdős–Rényi for larger $\langle k\rangle$.  The Erdős–Rényi is found to be the one with the worst overall quality, differing only slightly from the Barabási-Albert behaviour.

\subsection{Concentrating the squeezing}
The overall quality $\mu$ is the quantity to optimize if we want to use the cluster in MBQC protocols. On the contrary, if we want to use just two nodes to perform a quantum teleportation protocol, the best we can do is to concentrate the entanglement correlations in the two nodes: this corresponds to concentrating the squeezing on the nullifiers of the two chosen nodes.

Given a set of squeezing values for the input state  we use the protocol to concentrate the squeezing on the nullifiers of two nodes $n_1$ and $n_2$, using the fitness function $f(\Delta^2\hat{\delta}_i, n_1, n_2)=\sum_{i} A_i(n_1, n_2) \Delta^2\bar{\delta_i}$, where $n_1$ and $n_2$ are two given modes,. We set  $A_i(n_1, n_2)=10^5$ if $i=n_1, n_2$ and $A_i(n_1, n_2)=1$ otherwise. We see that the squeezing of the nullifiers is indeed concentrated on the two desired nodes, by reaching (and never exceeding) the highest squeezing values provided on the set of input states.

In this case we will indicate as $\mu_j$ the mean of the nodes of the graph $j$ that are not concerned with the teleportation protocol and with $\mu$ the mean of the $\mu_j$ values as follows:

\begin{equation}
\mu_j=\frac{1}{46}\sum_{i\neq n_1,n_2}\Delta^2\bar{\delta}_{i,j}\text{  }; \quad \mu=\langle \mu_j \rangle
\label{muall}
\end{equation}
where $\Delta^2\bar{\delta}_{i,j}$ identifies variance of the $i$-th nullifier of the $j$-th graph. $\mu_{n_{1}/n_{2}}$ denotes the mean of the set of values $\Delta^2\bar{\delta}_{n_{1}/n_{2},j}$, where $n_1$ and $n_2$ are two given nodes on which we chose to perform the teleportation.

As an example, in Table~\ref{BAtabletel} we show that for a 48-node Barabási-Albert model the squeezing on the two selected nodes $n_1$ and $n_2$ takes indeed the highest value provided on the input list of Fig.~\ref{listsqzval}. The same results hold for the Erdős–Rényi model and the Watts-Strogatz model.

\begin{table}
\centering
\begin{ruledtabular}
\begin{tabular}{cccc}
$m_{BA}$		&  $\mu_{12}$  (dB)		&$\mu_{13}$ (dB) 	 &$\mu$  (dB)	\\
\colrule
1			&-6.51		 	&-6.51			 	&-4.61		\\
5			&-6.51			&-6.51				&-5.48		\\
10			&-6.51			&-6.51				&-5.76		\\
20		&-6.51			&-6.51				&-6.10		\\
47			&-6.51			&-6.51				&-6.32		\\

\end{tabular}	\end{ruledtabular}
\caption{Means $\mu_{12}$, $\mu_{13}$ and $\mu$ of the nullifiers of the nodes 12 and 13 and of the value $\mu_j$ of Eq.~\eqref{muall} evaluated on $N=100$ Barabási-Albert graphs with different parameter $m_{BA}$, optimized using the function $f(\Delta^2\hat{\delta}_i)=\sum_{i} A_i\Delta^2\bar{\delta_i}$, where $A_i=10^5$ if $i=12, 13$ and $A_i=1$ otherwise.}\label{BAtabletel}
\end{table}

\subsection{Creating a quantum channel between nodes by manipulating existing networks}
In the previous section we have seen how to optimize the generation of complex cluster when we have multimode squeezed vacuum modes and we perform linear optics transformation. As already said, several experimental setups demonstrated the ability to deterministically generate large clusters following this approach; therefore, they can be used to generate the complex clusters presented above. \\
The generated cluster can then be distributed between different parties, i.e. different optical modes corresponding to different nodes of the network can be sent to different players, which can eventually use the entanglement correlations between the shared nodes for quantum communication protocols. According to the particular task, it may be necessary to reshape the entanglement correlations among the set of nodes.
We consider here the simplest case: the protocol wants to establish a maximally entangled state between two arbitrary nodes of a network in order to use it for quantum teleportation. The two-nodes entangled state is a two mode-squeezed state also called EPR, as it is the approximation of the entangled state used by Einstein, Poldolsky and Rosen in their famous paper in 1935~\cite{EPRpaper}.

The task of generating an entanglement link between chosen nodes corresponds to what is called quantum routing. As already said the procedure we follow  is well suited to CV entangled networks as it is relatively easy to deterministically generate the resources and then reshape them, while in the DV case the generation of optical entangled networks is costly and not deterministic so that the best procedure consists in routing the right entanglement connection at the beginning.

In the following we allow only for the easiest operations, i.e. linear optics transformations, for the reshaping. These can be global, when they operate on all the nodes at the same time, or local, when they act on a subset of nodes.

The case of a global transformation is somewhat trivial, as, if we are provided with a cluster $A$, implemented with a linear optics transformation $S_A$ acting on a set of squeezed input states, it is always possible to find the transformation that leads us to the cluster $B$. This transformation is simply $S=S_B\cdot S_A^{-1}$, where $S_B$ is the linear optics transformation that we should perform on the same set of input states to build the cluster $B$.

\begin{figure}
\centering
\includegraphics[width=\columnwidth]{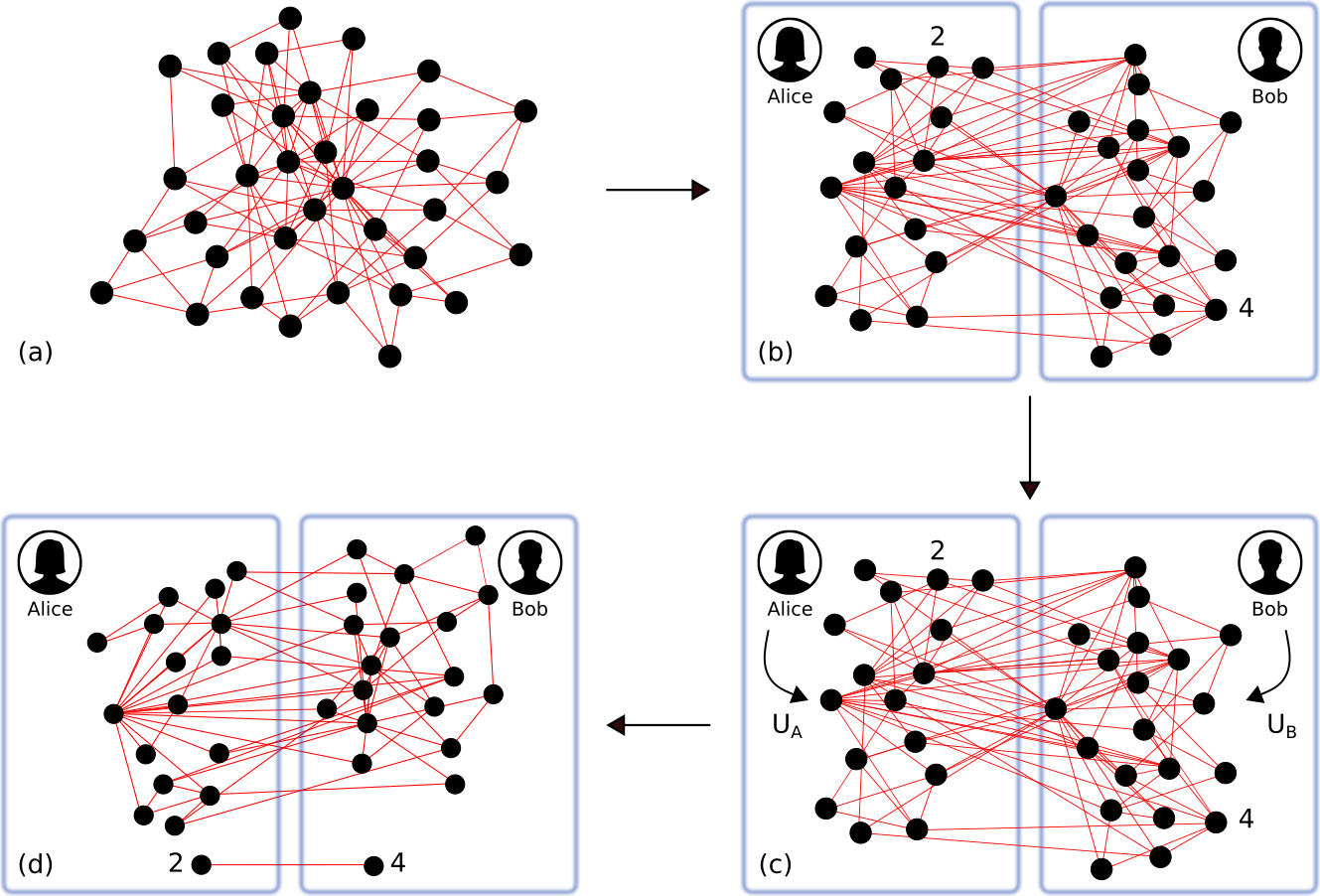}
\caption{(a) A quantum network is created. (b) The resource is distributed to two spatially separated teams, Alice and Bob. (c) Alice performs a linear optics operation $U_A$ on her set of nodes and Bob performs a linear optics operation $U_B$ on his set of nodes to create a quantum channel out of two given nodes. (d) The quantum channel is established. }
\label{division}
\end{figure}

We now consider a more interesting scenario: the modes of the cluster are distributed to two spatially separated parties, such that each party is allowed to perform local linear optics transformations only on its set of nodes, as shown in Fig.~\ref{division}. We then want to check which cluster shapes can be reconfigured in order to get a teleportation channel between two arbitrary nodes. 
A solution to this problem has already been found if we allow for more general symplectic transformations and weighted graphs~\cite{teamwork}.

Let’s say $n$ and $p$ are the number of nodes of party A (Alice) and B (Bob) respectively. We want to act with a linear optics transformation $U_A$ locally on the $n$ modes and with a linear optics transformation $U_B$ locally on the $p$ modes. The transformation acting on the whole set of quadratures of the modes then reads

\begin{equation}
\label{ggg}
S=\begin{pmatrix} \mathrm{Re}(U_A) & 0 & -\mathrm{Im}(U_A) & 0\\0&\mathrm{Re}(U_B)&0&-\mathrm{Im}(U_B)\\ \mathrm{Im}(U_A)&0& \mathrm{Re}(U_A)&0\\0&\mathrm{Im}(U_B)&0&\mathrm{Re}(U_B) \end{pmatrix}
\end{equation}
where $U_A$ and $U_B$ are two unitary matrices parametrized respectively by $n^2$ and $p^2$ parameters~\cite{Menicucci1}. A method for the generation of numerically random unitary matrices is presented in~\cite{Random}. If we define $\sigma_{1}$ as the covariance matrix of the cluster we are given and $\sigma_{2}$ as the covariance matrix of the cluster we obtain after the transformation, it holds 

\begin{equation}
\sigma_{2}=S \sigma_{1} S^T
\end{equation}
where $S$ is defined in Eq.~\eqref{ggg}. Our goal is to find the two matrices $U_A$ and $U_B$, whose real and imaginary parts define $S$, such that $\sigma_{2}$ is of the desired form. In our case we want $\sigma_{2}$ to be such that, for two given nodes, we get an EPR channel. In this case, it is hard to find an analytical solution to the problem, so we use a Derandomized Evolution Strategy (DES) algorithm to explore the parameter space~\cite{Jon1}.

If $n_1$ and $n_2$ are the nodes out of which we want to obtain an EPR state, we can define

\begin{gather}
\sigma_{2, red}(n_1, n_2)=\begin{pmatrix} \sigma_{2, n_1}\\\sigma_{2, n_2}\\\sigma_{2, n_2+n+p}\\\sigma_{2, n_2+n+p} \end{pmatrix}
\end{gather}
as the reduced $4\times 2(n+p)$ matrix obtained by selecting only the rows of $\sigma_{2}$ we want to set as a quantum channel. We thus define the $4\times 2(n+p)$ matrix $\sigma_{channel}$ as the matrix with null entries except for

\begin{gather}
\sigma_{1,n_1}=\sigma_{2,n_2}=\sigma_{3,n_1+n+p}=\sigma_{4,n_2+n+p}=\lambda\\
\sigma_{1,n_2+p+n}=\sigma_{2,n_1+p+n}=\sigma_{3,n_2}=\sigma_{4,n_1}=\mu
\end{gather}
where $\mu$ and $\lambda$ are numerical values that we fix according to the squeezing we want to attain on the modes of the quantum channel. This squeezing cannot exceed the squeezing of the input states of the cluster. For simplicity, we worked with an equal value of squeezing, both for the input squeezed states used to implement the cluster and for the squeezing chosen for the quantum channel. We will search for the minimal value of the function

\begin{equation}
\label{foptlocal}
f_{opt}=|| \sigma_{channel}-\sigma_{2, red}(n_1, n_2) ||
\end{equation}

where $||\cdot||$ indicates the Frobenius norm. 

\begin{figure}[h]
\captionsetup[subfigure]{labelformat=empty,width=0.3\textwidth, margin={0cm,0.1cm}}
\centering
\subfloat[][\emph{6-mode Grid}]  
 {\includegraphics[scale=0.10]{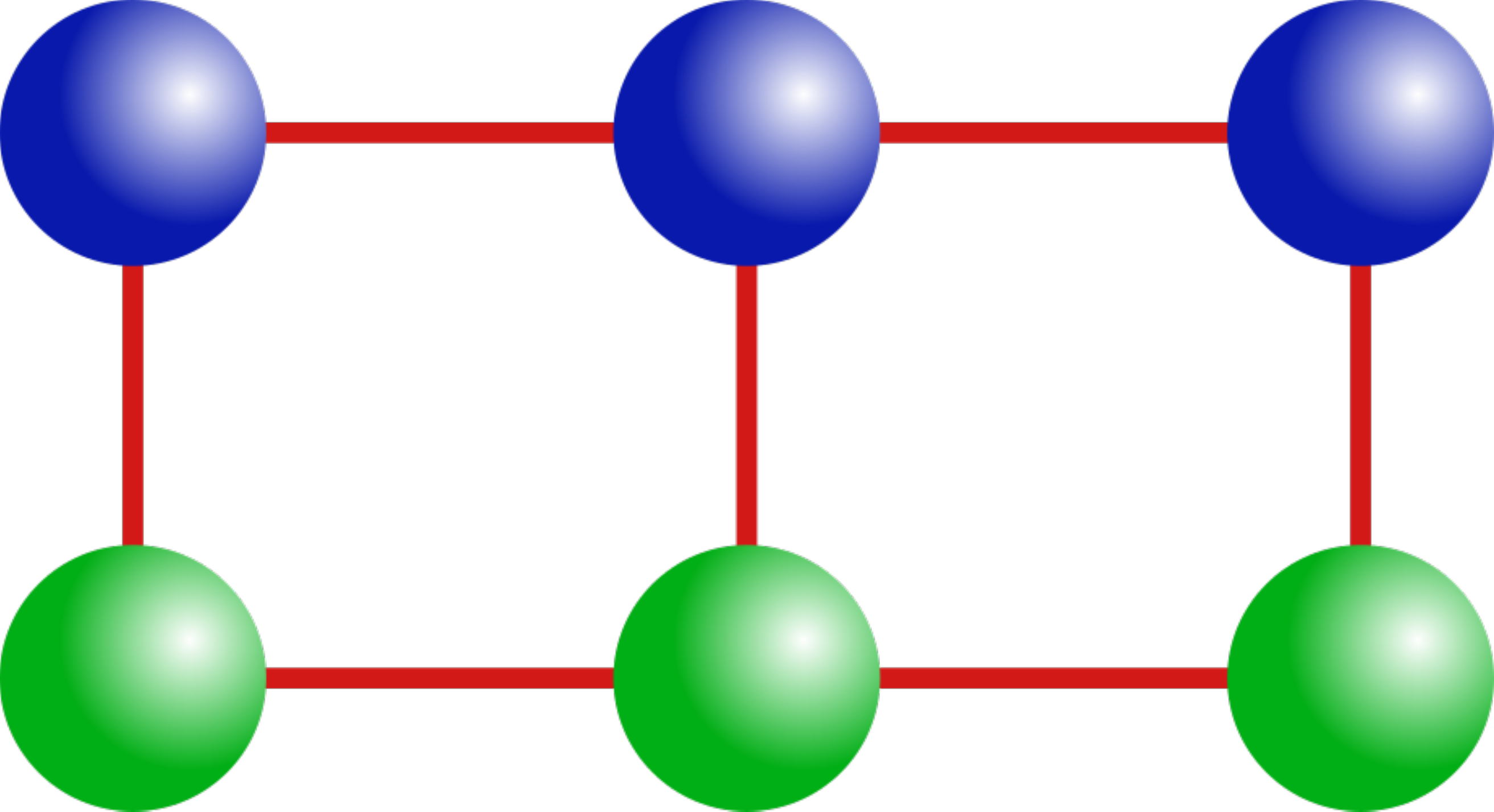}} \quad\quad\quad
 \subfloat[][\emph{8-mode Grid}]  
{\includegraphics[scale=0.10]{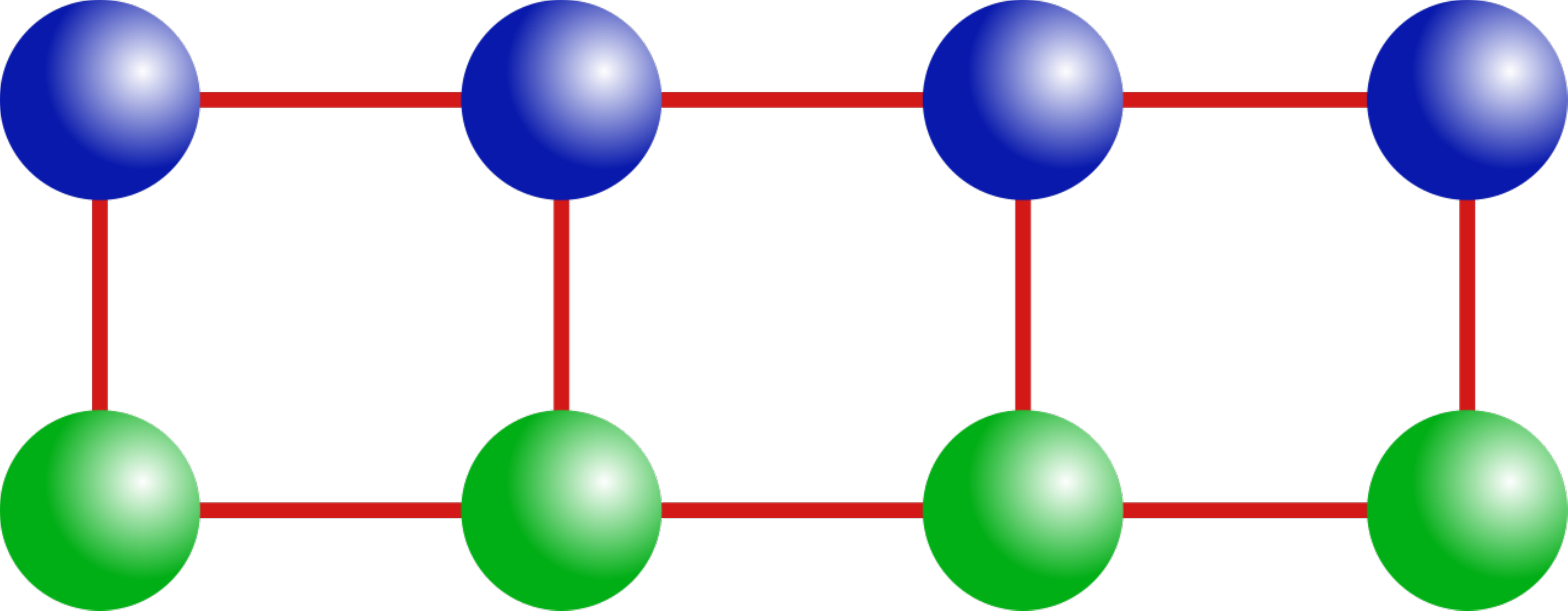}}\\
\subfloat[][\emph{10-mode Grid}]  
{\includegraphics[scale=0.10]{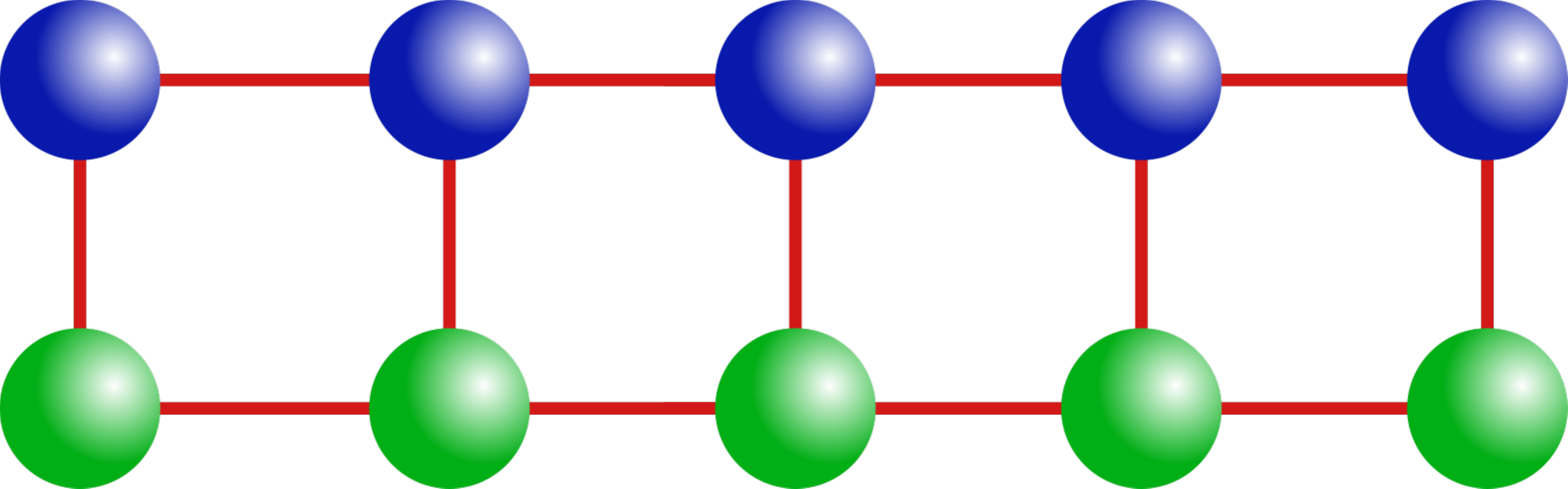} \label{grid}}\quad\quad\quad
\subfloat[][\emph{Fully-connected}]  
 {\includegraphics[scale=0.10]{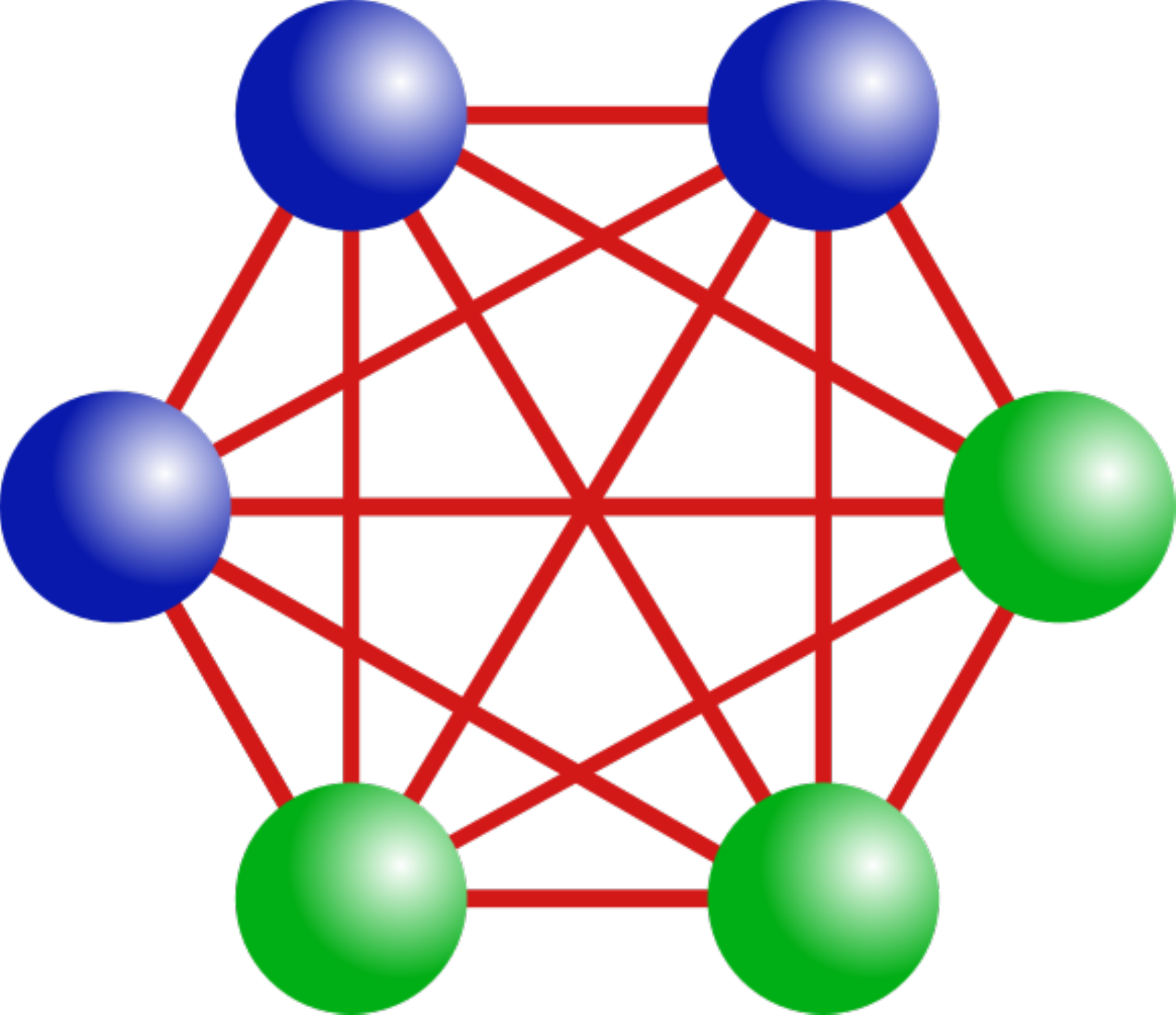}\label{fully}} \\
\subfloat[][\emph{Graph ``X''}]  
 {\includegraphics[scale=0.10]{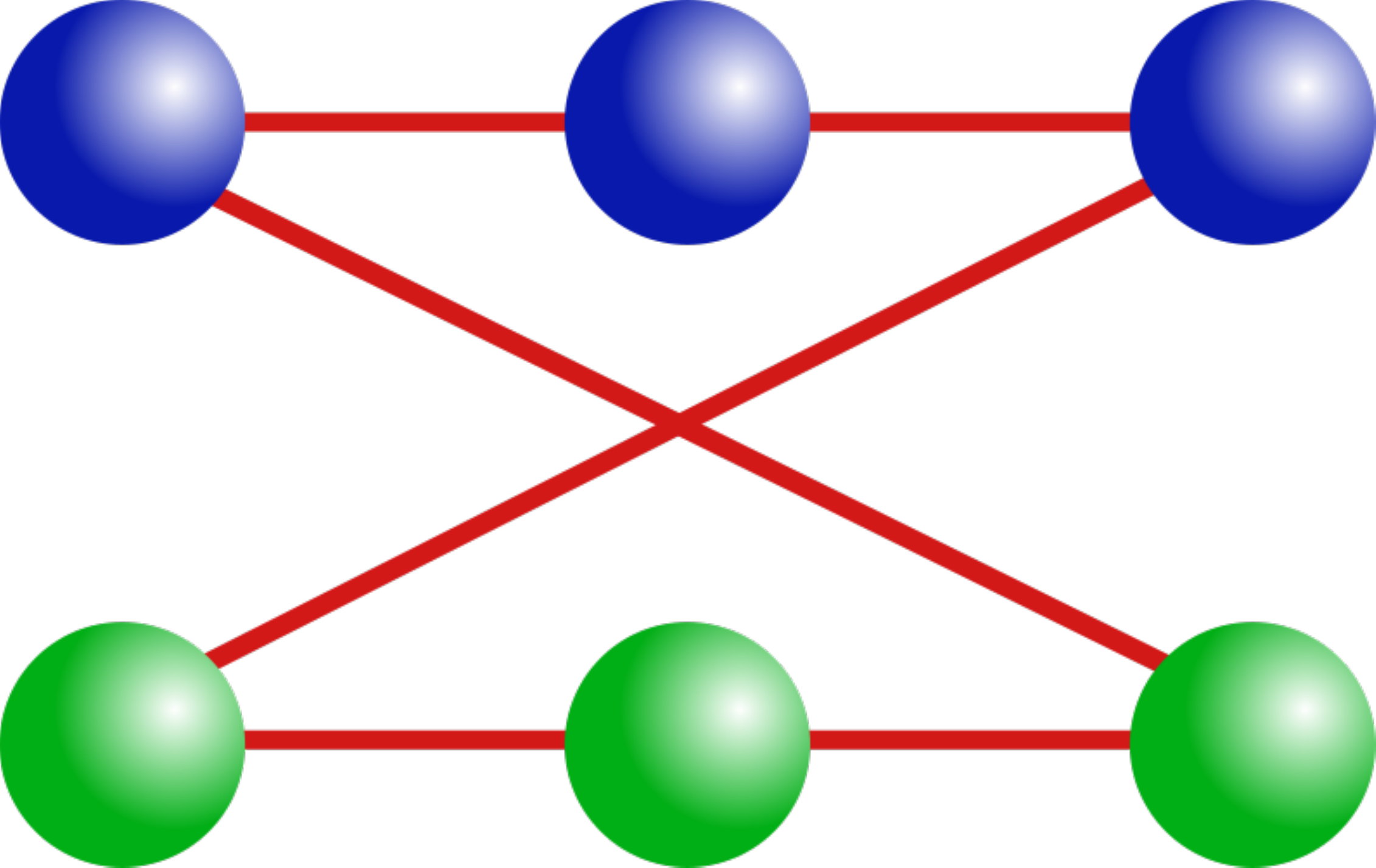}\label{yes1}} \quad\quad\quad
 \subfloat[][\emph{Graph ``Y''}]  
{\includegraphics[scale=0.10]{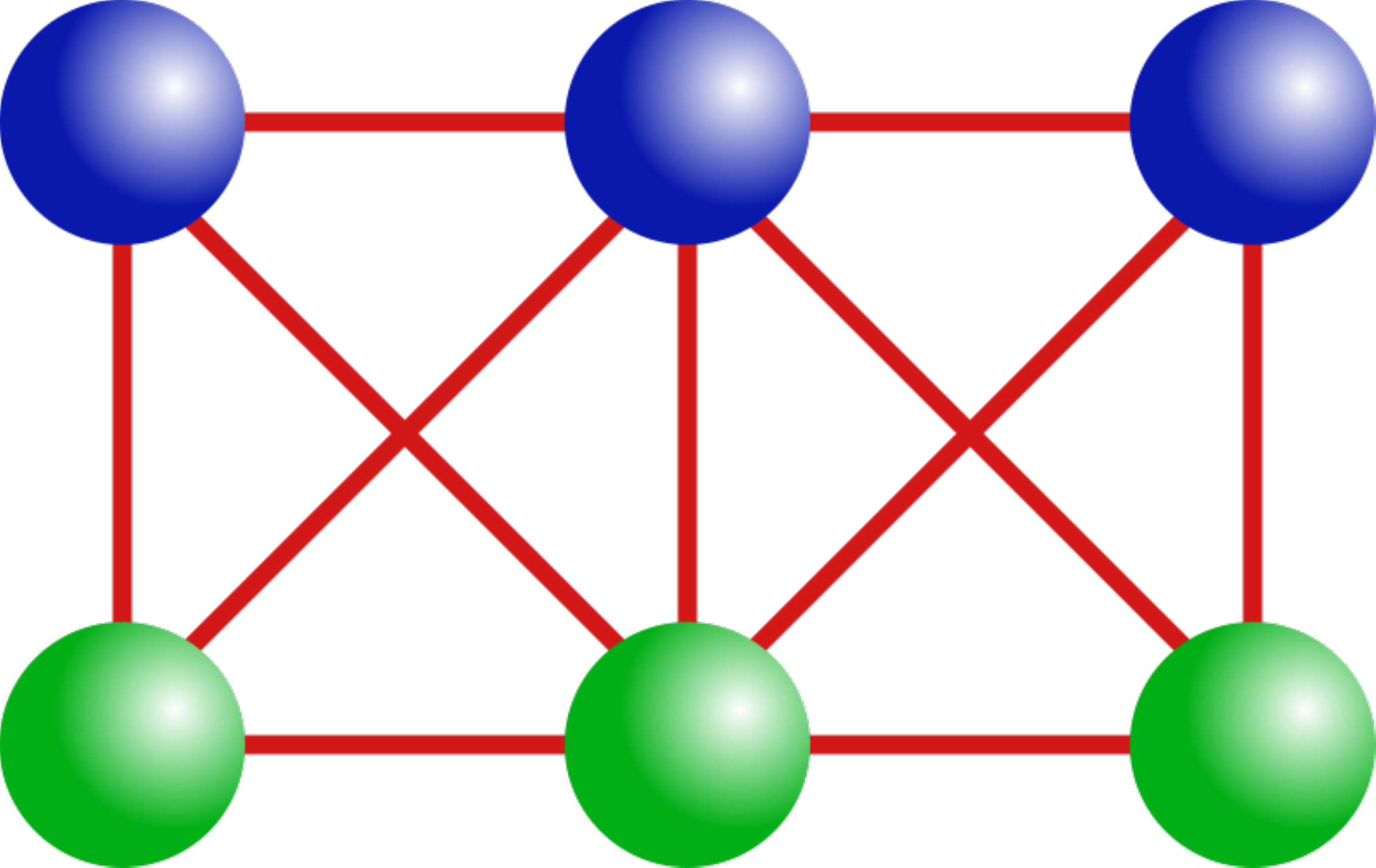}\label{yes2}}\\
\subfloat[][\emph{Graph ``Z''}]  
 {\includegraphics[scale=0.10]{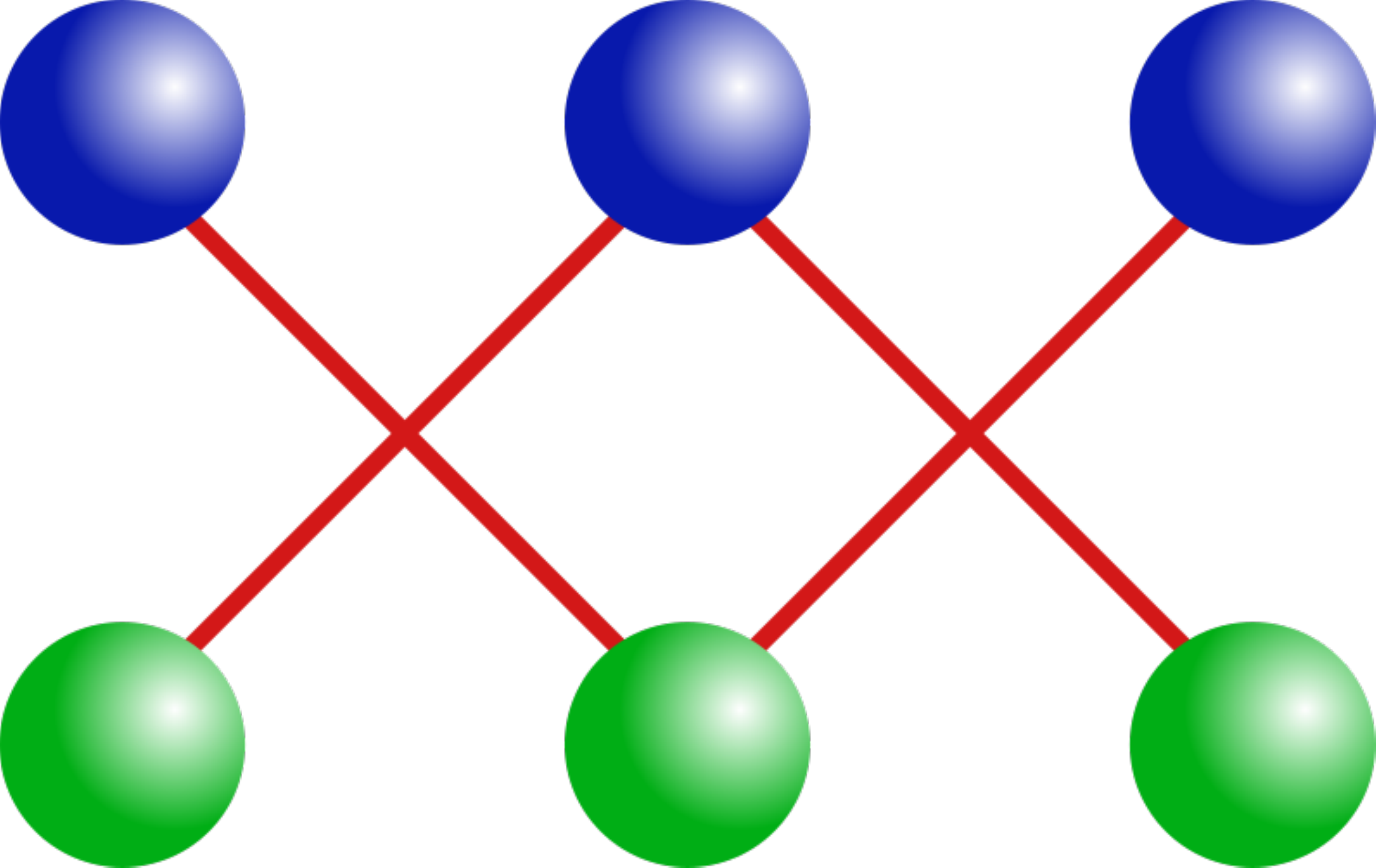}\label{no1}} \quad\quad\quad
 \subfloat[][\emph{Dual-rail}]  
{\includegraphics[scale=0.10]{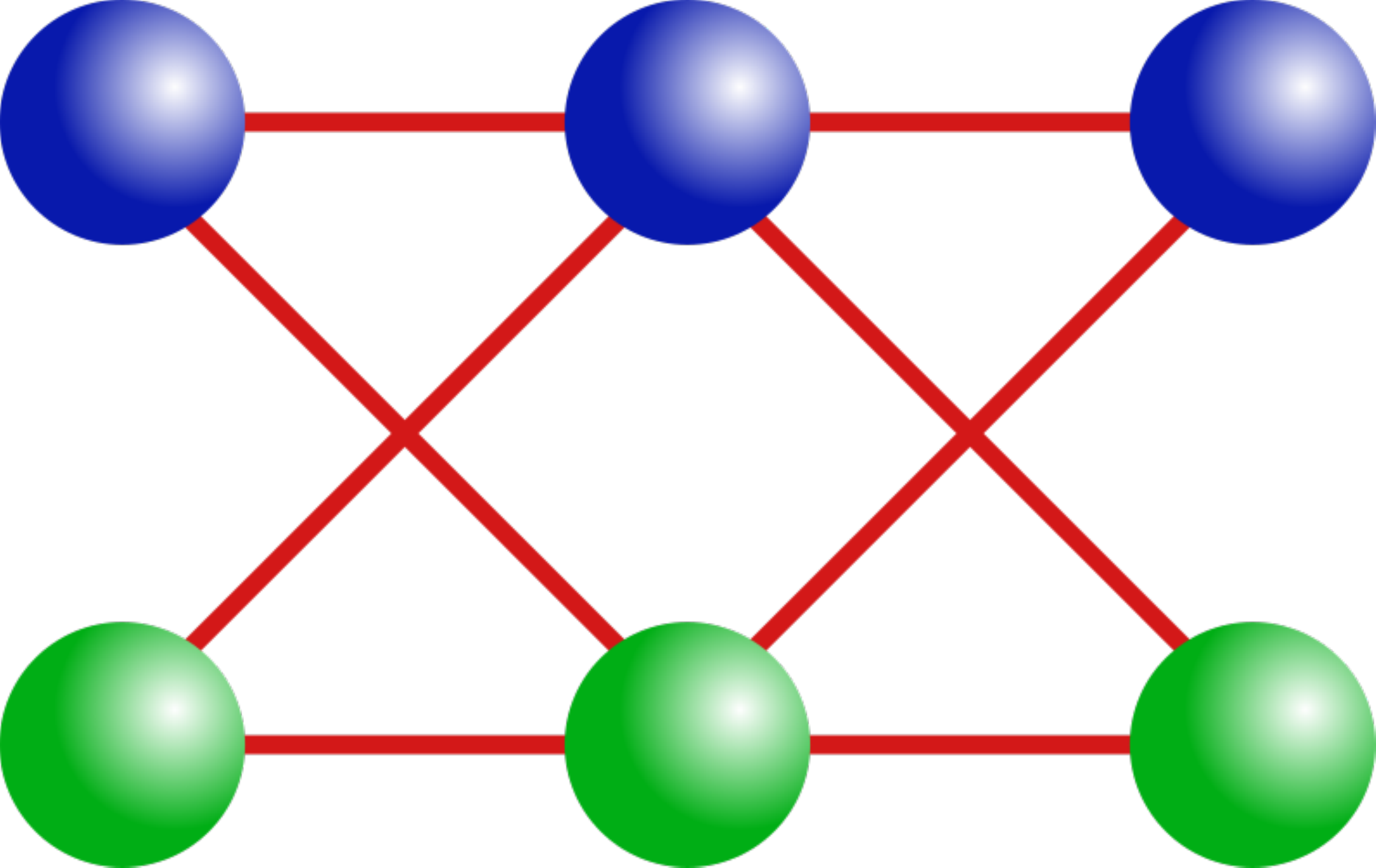}\label{no2} }
\caption{Graphs analysed with the aim of creating an EPR channel between Alice (green) and Bob (blue) (or eventually between nodes of the same team).}
\label{structures}
\end{figure}

In Table~\ref{tab:results} we show preliminary results on different graphs with a restricted number of nodes, shown in Fig.~\ref{structures}. For the 6-mode and 10-mode ``grid'' graphs and the graphs ``X'' and ``Y'' a result can be found for the creation of a quantum channel between Alice and Bob. For these structures, however, it was not possible to find a solution for the creation of a quantum channel between nodes of the same team. The opposite stands for the fully-connected cluster, for which it was possible to create a channel between nodes of the same team but not between nodes of different teams. Lastly, for the graph ``Z'', which represents two 3-mode cluster states distributed to the two parties, for the dual-rail and for the 8-mode ``grid'' graph, no solution was found. As an example, the results of the fully-connected graph and of the 6-mode ``grid'' graph of Fig.~\ref{grid} are shown in Appendix.

\begin{table}[h]
\centering
\begin{ruledtabular}
\begin{tabular}{ccc}
Graph					&Between A and B			&Same team\\
\colrule
6-node grid			&Yes								&No\\
8-node grid			&No								&No\\
10-node grid			&Yes								&No\\
Fully-connected	&No									&Yes\\
``X''						&Yes								&No\\
``Y''						&Yes									&No\\
``Z''						&No									&No\\
Dual-rail				&No									&No\\
\end{tabular}	\end{ruledtabular}
\caption{Results on the possibility to create a quantum communication channel for the graphs of Fig.~\ref{structures}, between nodes belonging to different teams A and B and between nodes belonging to the same team.}
\label{tab:results}
\end{table}

\section{Discussion} 

We have shown that for CV cluster states with complex graphical representation the quality, measured as the mean of the variance of the nullifiers, can be better optimized when the quantity of entanglement links increases. We have analyzed in the quantum regime different complex shapes corresponding to different models of real-world networks. In the Barabási-Albert and Erdős–Rényi model, regardless of the topology, clusters with a similar average degree $\langle k \rangle$ have a comparable overall quality. The average degree $\langle k \rangle$ in complex graphs could thus be used as a benchmark for the quality of the state implemented with the optimization protocol. Moreover, analyzing ``small-world networks'', that evolve from a regular network to a random network as their characteristic parameter $p_{WS}$ increases, we found that randomness in the structure is detrimental to the quality of the state. The optimization procedure can be also used to concentrate the entanglement between two given nodes. Global optimizations of cluster states via linear optics unitaries can always be realized via analytical procedures. \\
On the contrary, if the cluster is distributed in different locations and the players want to reshape it via local operations (quantum routing) with the aim of performing quantum communication protocols, we need to use a numerical procedure in order to deal with the larger number of constraints. \\
In this case we have used a Derandomized Evolutionary Strategy (DES) algorithm with a suitable fitness function, the one of eq ~\eqref{foptlocal}, in order to find solutions to generate an EPR state between two arbitrary nodes of a network shared between two parties, which can perform only local linear optics operations. 

We have studied small networks and we have found that it is possible to create EPR channels between two nodes of the two different teams for some network shapes, while creating an EPR channel from two nodes of the same team has never been possible, except for the case of the fully connected network. So, except for the last case, it has is never been possible, by using only local linear optics operations, to disconnect two nodes of one player from the ones of the other player.
It has to be stressed that in the cases where no solution is found, we cannot conclude with certainty that the solution does not exist, as the DES algorithm can be stuck in a local minimum in the parameters space.
%

In future works we will investigate quantum routing operations when complex cluster are distributed between several parties by also adding, when linear optics operations are not sufficient, quadrature measurements. The measurement of the$q$ quadrature of a mode of the cluster allows in fact for node removal, while the measurement of $p$ is useful in wire shortening~\cite{Gu}: both can be used in order to cut the residual edges after the optimization via linear optics operations.

\section{Materials and Methods}
To implement the networks and carry out the data analysis we used Wolfram Mathematica Version 11.3. In particular, the Barabási-Albert, Erdős–Rényi and Watts-Strogatz model are already embedded in the software. Wolfram Mathematica has been used also to implement the DES \textit{($\mu$-$\lambda$) iso-CMA algorithm} presented in~\cite{Jon1}. The goal of a DES algorithm is the optimization of a given function $f(\mathbf{x})$, where $\mathbf{x}$ is a vector of parameters. We are free to choose our starting point $\mathbf{x}^{old}$ in the parameter landscape and we ``mutate'' it, generating $\lambda$ new points (\textit{offspring}) as
\begin{equation}
\mathbf{x}_k^{new}=\mathbf{x}^{old}+\Delta\mathbf{x}_k\text{, where }k=1, \dots, \lambda
\end{equation}

where $\Delta\mathbf{x}_k$ is drawn from a multivariate normal 
distribution  $\mathcal{N}(0, \sigma^2\mathbbold{1})$. 
The $\lambda$ new points are then evaluated with respect to the chosen fitness function $f$ and sorted. The $\mu$ ``mutants'' that provide the best result are chosen to generate a new parent as
\begin{equation}
\mathbf{x}^{new}=\sum_{k}w_k\mathbf{x}_k^{new}
\end{equation}
where $\sum_k w_k=1$. The procedure is then iterated, setting $\mathbf{x}^{new}$ as the new starting point. A learning component is provided by updating, at each generation, the global step-size $\sigma$.
As already said above, in our case we want to reach a specific value $f_{opt}\sim 0$ of a non-negative fitness function. However, it is never guaranteed that the DES procedure finds the global extremum. 

\begin{acknowledgments}
The authors acknowledge financial support from the European Research Council under the Consolidator Grant COQCOoN (Grant No. 820079). The authors also thanks F. Arzani, N. Treps and M. Walschaers for useful discussions.
\end{acknowledgments}

\appendix
\section{}

\begin{figure}[h]
\centering
\subfloat[][\emph{}]  
 {\includegraphics[scale=0.12]{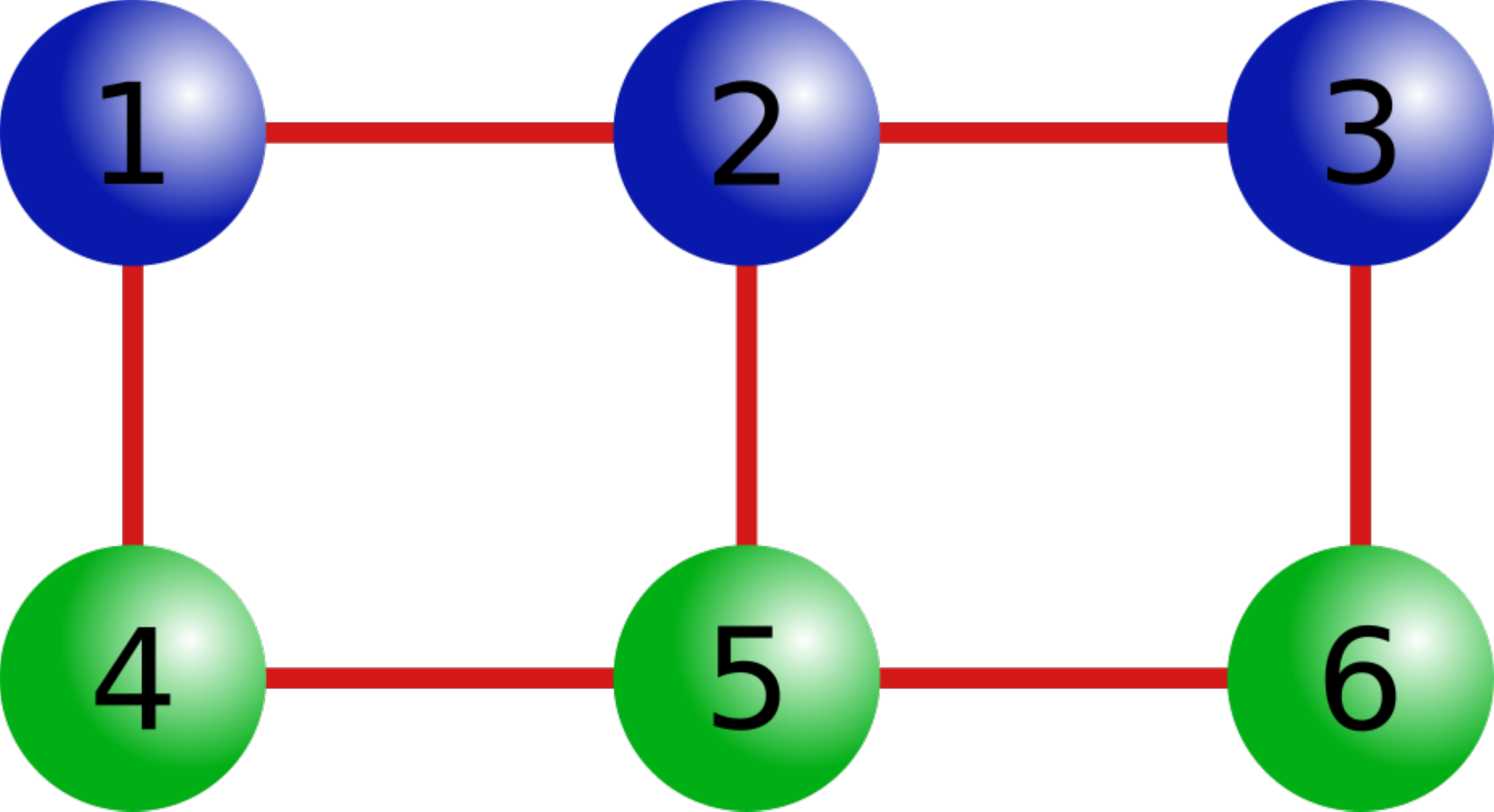} \label{gridnumb} }\quad
\subfloat[][\emph{}]  
{\includegraphics[scale=0.12]{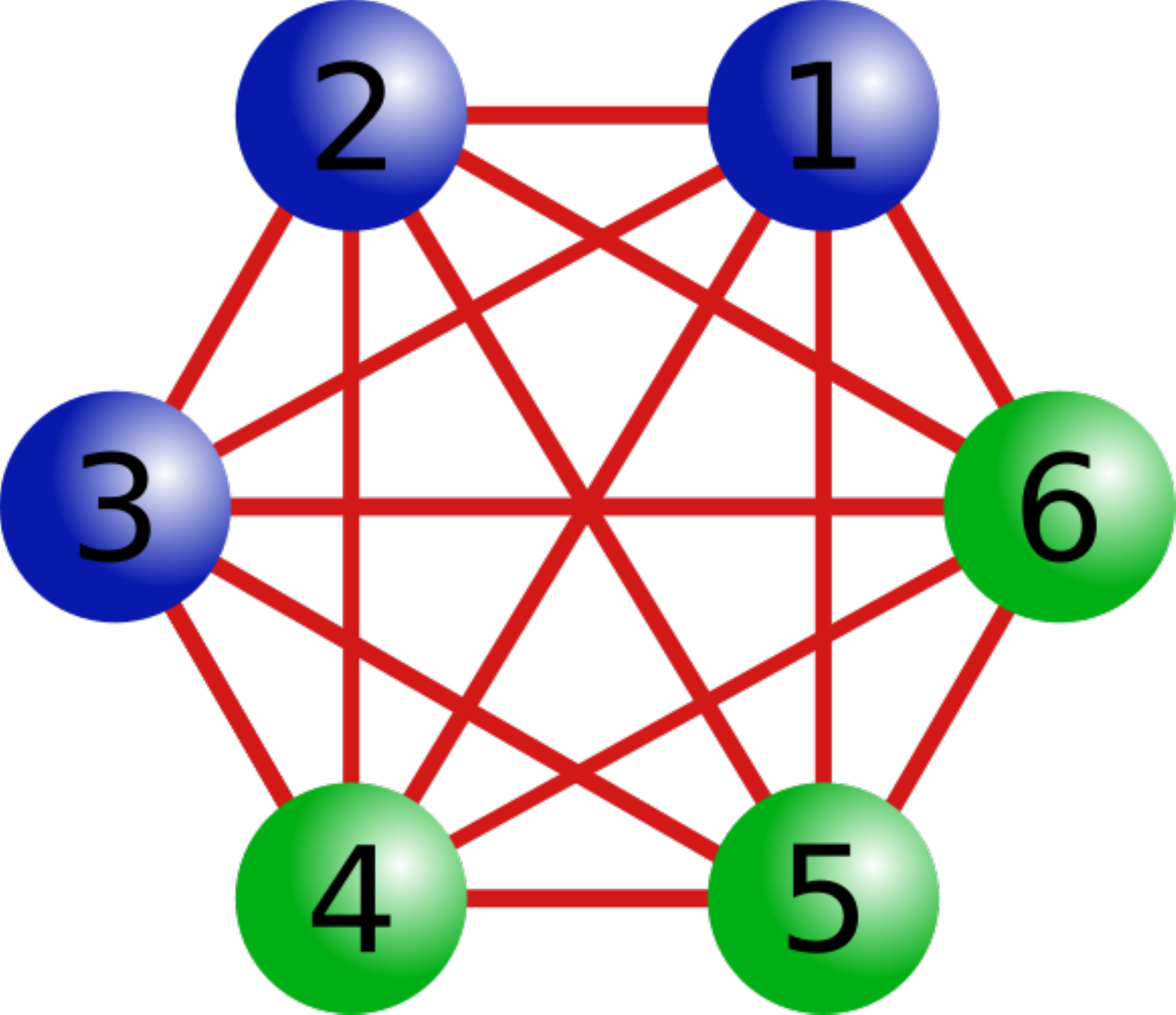}\label{fully2} }
\caption{Graphs whose results for the creation of an EPR channel between Alice (green) and Bob (blue) are shown in this below.}
\label{appendixpic}
\end{figure}
We will present here the result for the creation of a quantum channel out of two given nodes for the graphs shown in Fig.~\ref{appendixpic}.

For the 6-mode ``grid'' graph of Fig.~\ref{gridnumb} the suitable unitary matrices $U_A$ and $U_B$ (see Eq.~\eqref{ggg}) that we need for the creation of a quantum channel out of the nodes 1 and 4 are

\begin{gather*}
\mathrm{Re}(U_A)=\left(
\begin{array}{ccc}
 -0.564055 			& \mathcal{O}(10^{-16})			& 0.564055 \\
 0.250315 			& \mathcal{O}(10^{-16}) 			& 0.250315 \\
 \mathcal{O}(10^{-16}) 	& -0.277133 					& \mathcal{O}(10^{-16}) \\
\end{array}
\right)\\
\mathrm{Im}(U_A)=\left(
\begin{array}{ccc}
 -0.426429 			&\mathcal{O}(10^{-16}) 			& 0.426429 \\
 -0.661319 			&\mathcal{O}(10^{-17}) 			& -0.661319 \\
\mathcal{O}(10^{-16}) 	& -0.960831 					&\mathcal{O}(10^{-16}) \\
\end{array}
\right)\\
\mathrm{Re}(U_B)=\left(
\begin{array}{ccc}
 -0.564055 			&\mathcal{O}(10^{-17}) 			&0.564055 \\
 -0.449914 			&\mathcal{O}(10^{-17}) 			& -0.449914 \\
\mathcal{O}(10^{-18}) 	& -0.993175 					&\mathcal{O}(10^{-17}) \\
\end{array}
\right)\\
\mathrm{Im}(U_B)=\left(
\begin{array}{ccc}
 0.426429 			&\mathcal{O}(10^{-17}) 			& -0.426429 \\
 -0.545507 			& \mathcal{O}(10^{-18}) 			& -0.545507 \\
\mathcal{O}(10^{-18}) 	& -0.116635 					&\mathcal{O}(10^{-17})\\
\end{array}
\right)
\end{gather*}
Using these transformations, the correlations between 1 and 4 and the other nodes are at most of the order of $10^{-15}$.

For the ``fully connected'' graph of Fig.~\ref{fully2} the suitable unitary matrices $U_A$ and $U_B$ (see Eq.~\eqref{ggg}) that we need for the creation of a quantum channel out of the nodes 1 and 2 are

\begin{gather*}
\mathrm{Re}(U_A)=\left(
\begin{array}{ccc}
 0.56149 		& -0.397134 	& -0.164356 \\
 -0.56149 		& 0.397134 	& 0.164356 \\
 0.408248 	& 0.408248 	& 0.408248 \\
\end{array}
\right)\\
\mathrm{Im}(U_A)=\left(
\begin{array}{ccc}
 -0.134394 & -0.419068 & 0.553462 \\
 -0.134394 & -0.419068 & 0.553462 \\
 -0.408248 & -0.408248 & -0.408248 \\
\end{array}
\right)\\
\mathrm{Re}(U_B)=\left(
\begin{array}{ccc}
 0.447715 & 0.293987 & 0.104392 \\
 0.436176 & 0.614297 & -0.472751 \\
 0.124502 & 0.361237 & 0.860632 \\
\end{array}
\right)\\
\mathrm{Im}(U_B)=\left(
\begin{array}{ccc}
 -0.706639 & 0.450256 & -0.0124474 \\
 0.201155 & -0.408378 & 0.0407497 \\
 0.232375 & -0.190305 & 0.152002 \\
\end{array}
\right)
\end{gather*}
Using these transformations, the correlations between 1 and 2 and the other nodes are at most of the order of $10^{-15}$.

\nocite{*}

\bibliography{apssamp.bib}

\end{document}